\documentclass{DISproc}
\def\pb{\,{\rm pb}}
\begin{document}
\title{Timelike Compton Scattering - New Theoretical Results and Experimental Possibilities}

\author{{\slshape B. Pire$^1$, L. Szymanowski$^2$, Jakub Wagner$^2$}\\[1ex]
$^1$ CPHT, {\'E}cole Polytechnique, CNRS, 91128 Palaiseau, France\\
$^2$ National Center for Nuclear Research (NCBJ), Warsaw, Poland
 }

\contribID{xy}

\doi  

\maketitle

\begin{abstract}
 We review recent progress in the study of timelike Compton scattering (TCS), the crossed process of deeply virtual Compton scattering. We emphasize the need to include NLO corrections to any phenomenological program to extract Generalized Parton Distributions (GPDs) from near future experimental data. We  point out that TCS at high energy should be available through a study of ultraperipheral collisions at  RHIC and LHC, opening a window on quark and gluon GPDs at very small skewness.
\end{abstract}

\section{Intoduction}
Almost two decades after its first stages~\cite{Muller:1994fv}, the study of deeply virtual Compton scattering (DVCS),
 i.e., $\gamma^* p \to \gamma p$, and more generally of hard exclusive reactions in a generalized Bjorken regime, has now entered a phase where many theoretical and experimental progresses can merge to enable a sensible extraction of  generalized parton
distributions (GPDs). 
Indeed, the measurement of GPDs should contribute in a decisive way to
our understanding of how quarks and gluons build hadrons~\cite{gpdrev}. In particular the transverse
location of quarks and gluons become experimentally measurable via the transverse momentum dependence of the GPDs \cite{Burk}.

Timelike Compton scattering (TCS) \cite{TCS}  $$\gamma(q) N(p) \to \gamma^*(q') N(p') \to l^-(k) l^+(k') N(p')$$
   at small $t = (p'-p)^2$ and large \emph{timelike} virtuality $(k+k')^2=q'^2 = Q^2$ of the final state
 dilepton, shares many features with its ``inverse'' process, DVCS. The Bjorken variable in the TCS case is $\tau = Q^2/s $
 with $s=(p+q)^2$. One also defines $\Delta = p' -p$  ($t= \Delta^2$) and the skewness variables 
$\xi  = - \frac{(q+q')^2}{2(p+p')\cdot (q+q')} \,\approx\,
           \frac{ - Q^2}{2s  - Q^2} $, 
$\eta = - \frac{(q-q')\cdot (q+q')}{(p+p')\cdot (q+q')} \,\approx\,
           \frac{ Q^2}{2s  - Q^2}$
where the approximations hold in  the extended Bjorken regime
where masses and $-t$ are small with respect of $Q^2$ ($s$ is always
larger than $Q^2$ ). 
\begin{figure}[htb]
  \centering
  \includegraphics[width=0.3\textwidth]{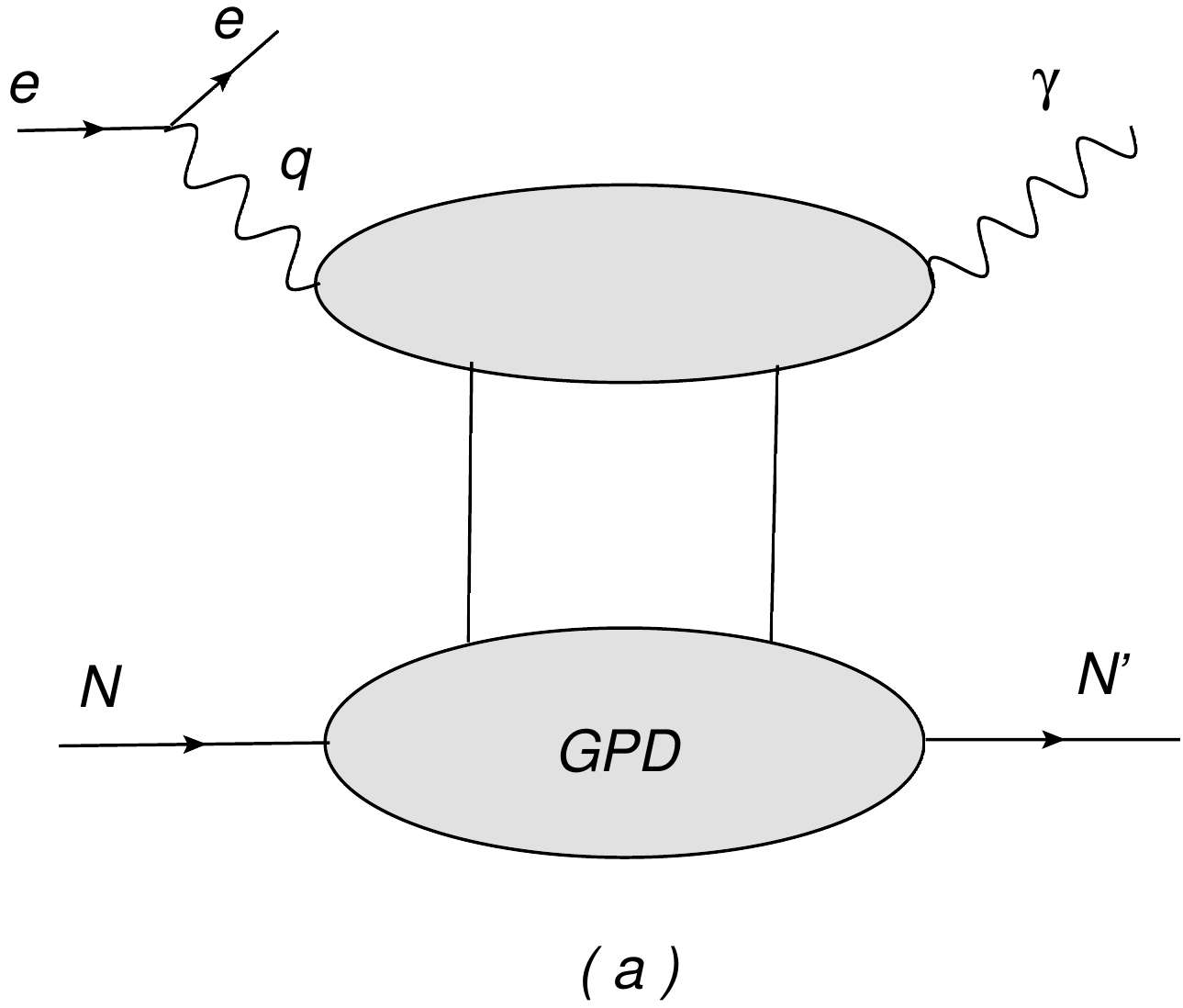}
  \includegraphics[width=0.3\textwidth]{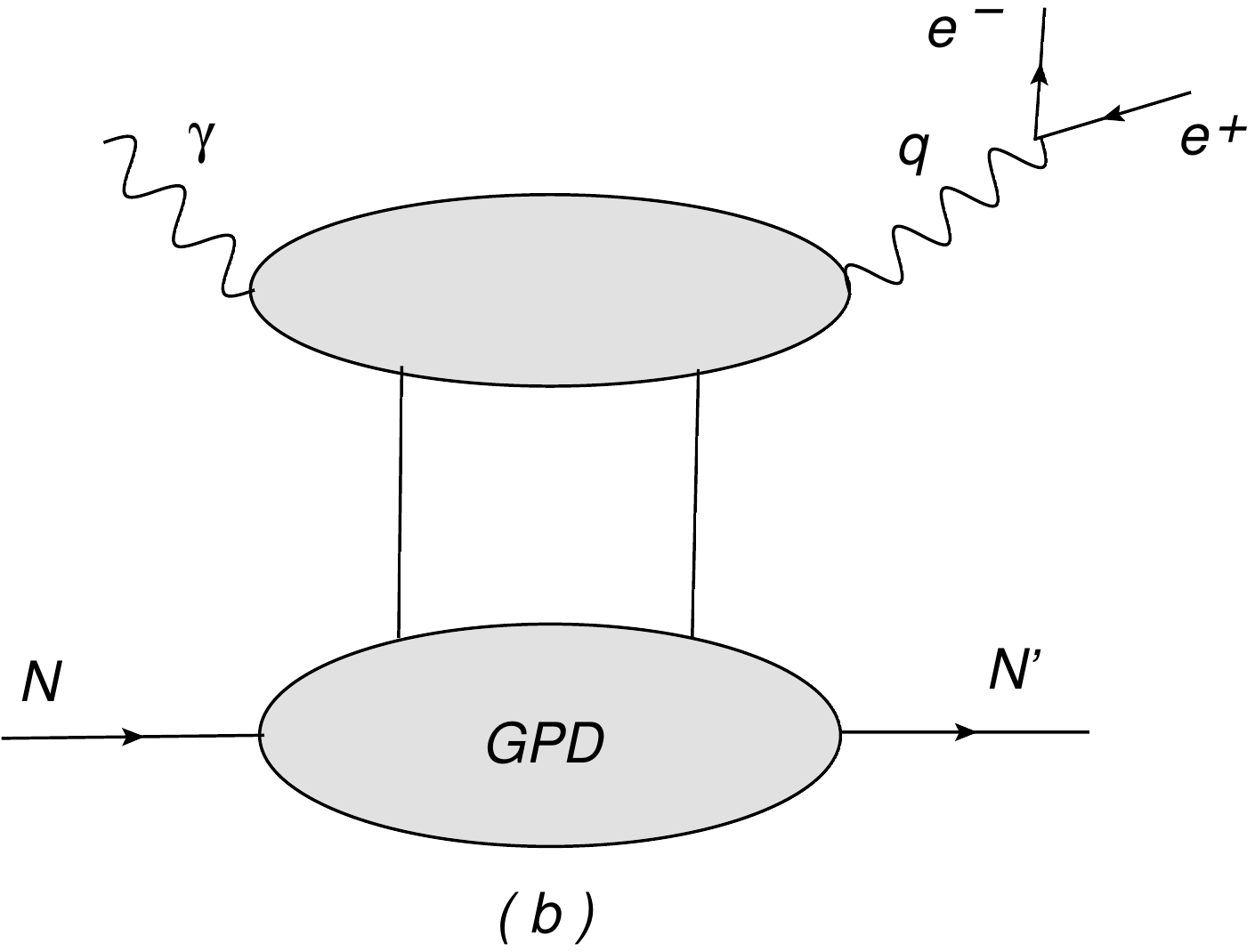}
  \caption{(a) Deeply Virtual Compton Scattering and (b) Timelike Compton Scattering}
  \label{Fig:DVCSTSC}
\end{figure}
\section{Basic properties and first experimental results}
In the region where the final photon virtuality is large, the Compton amplitude is given by the convolution of hard scattering coefficients, calculable in perturbation theory, and generalized parton distributions, which describe the nonperturbative physics of the process. The physical process where to observe TCS, is photoproduction of a heavy lepton pair, $$\gamma N \to \mu^+\!\mu^-\, N ~~~~~or~~~~~ \gamma N \to e^+\!e^-\, N\;.$$
A QED process, the Bethe-Heitler (BH)
mechanism  $\gamma(q)  \gamma^*(-\Delta)  \to l^-(k) l^+(k') $ contributes at the amplitude level. 
This latter process has a very peculiar angular dependence and overdominates the TCS process if
one blindly integrates over the final phase space. One may however choose kinematics where 
the amplitudes of the two processes are of the same order of magnitude, and  use specific observables sensitive to the interference of the two amplitudes. Since the amplitudes for the Compton and Bethe-Heitler
processes transform with opposite signs under reversal of the lepton
charge,  it is possible to project out the interference term through a clever use of the angular distribution of the lepton pair \cite{TCS}.

 First attempts to measure TCS, and to confront the theoretical predictions with data were already performed at JLab at 6 \GeV \cite{Rafayel}, and may serve as a feasibility test for a proposed experiment with higher energy after upgrade to 12 \GeV.

\section{TCS at next to leading order}
\label{sec:NLO}
After proper renormalization, the Compton scattering amplitude reads in its factorized form:
\begin{eqnarray}
\mathcal{A}^{\mu\nu} &=& -g_T^{\mu\nu}\int_{-1}^1 dx 
\left[
\sum_q^{n_F} T^q(x) F^q(x)+T^g(x) F^g(x)
\right] \nonumber \\
&+& i\epsilon_T ^{\mu\nu}\int_{-1}^1 dx 
\left[
\sum_q^{n_F} \tilde{T}^q(x) \tilde{F}^q(x)+\tilde{T}^g(x) \tilde{F}^g(x)
\right] \,,
\label{eq:factorizedamplitude}
\end{eqnarray}
where renormalized coefficient functions for the vector case are given by:
\begin{eqnarray}
T^q(x)&=& \left[ C_{0}^q(x) +C_1^q(x) +\ln\left(\frac{Q^2}{\mu^2_F}\right) \cdot C_{coll}^q(x)\right] - ( x \to -x )  \,,\nonumber\\
T^g(x) &=& \left[ C_1^g(x) +\ln\left(\frac{Q^2}{\mu^2_F}\right) \cdot C_{coll}^g(x)\right] +( x \to -x )
\,.
\label{eq:ceofficients}
\end{eqnarray} 
and similarily (but with different symmetry in $x$) for the axial quantities $\tilde{T}^q, \tilde{T}^g$. Results for TCS \cite{Pire:2011st} compare to the well-known DVCS results \cite{BelMueNieSch00}, through a simple relation \cite{MPSW}: 
\begin{eqnarray}
^{TCS}T(x) = \pm \left(^{DVCS}T(x) +  i \pi C_{coll}(x)\right)^* \,,
\label{eq:TCSvsDVCS}
\end{eqnarray}
where +(-) sign corresponds to vector (axial) case. 
\begin{figure}[htb]
  \centering
  \includegraphics[width=0.35\textwidth]{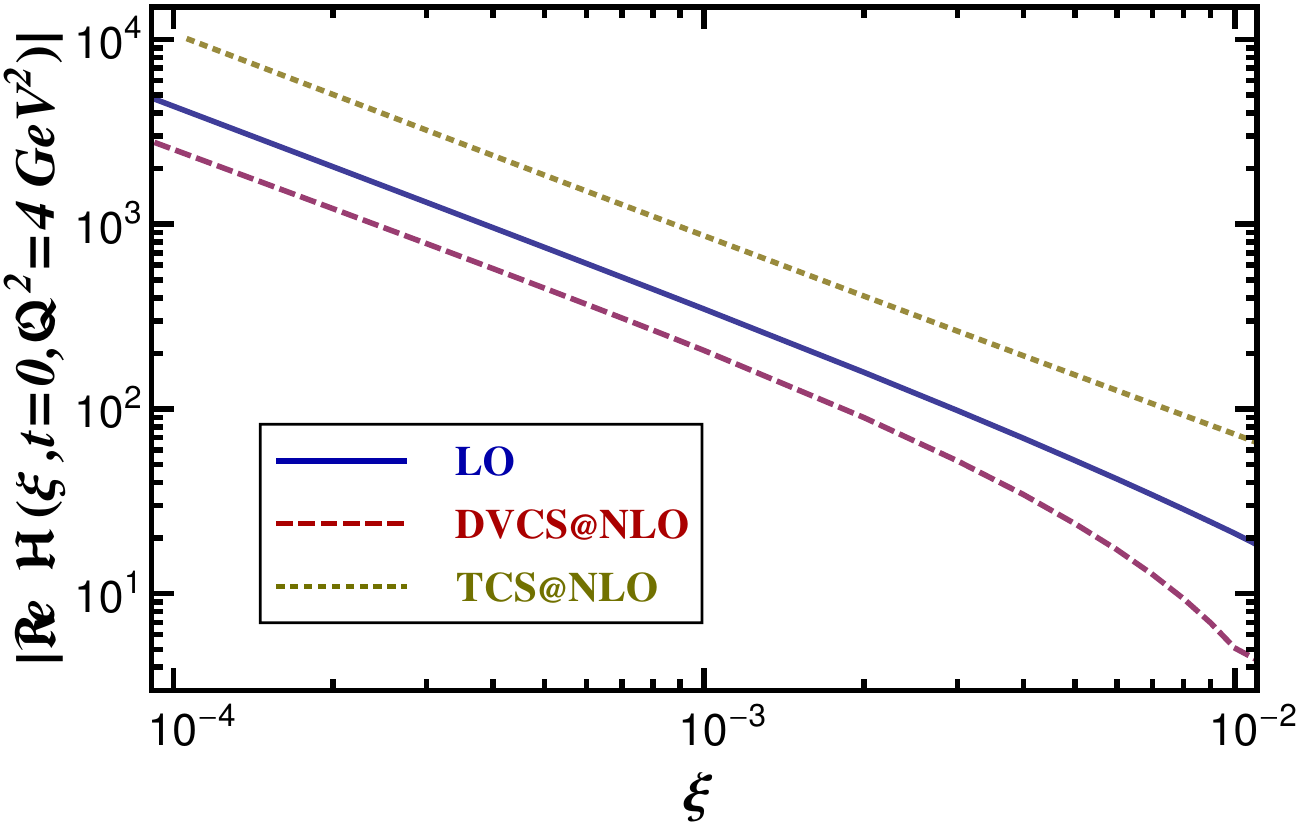}
  \caption{The real part of CFF $\mathcal{H}$ vs.~$\xi$ with $\mu^2=Q^2= 4 \textrm{~GeV}^2$  and $t=0$ at LO (solid) and NLO  for DVCS (dashed). For  TCS at NLO  its negative value is shown as dotted curve.}
  \label{fig:NLO}
\end{figure}
The NLO relation (\ref{eq:TCSvsDVCS}) tells us that if scaling violations are small, the timelike CFFs (TFFs) can be obtained from the spacelike ones by complex conjugations. Moreover, GPD model studies indicate that in the valence region, i.e., for $\xi \sim 0.2$,   CFFs  might only evolve mildly. This rather generic statement, which will be quantified by model studies \cite{Moutarde}, might be  tested in future (after 12GeV upgrade) Jefferson Lab experiments. On the other hand we expect huge  NLO corrections
to $\Re{\rm e}{^T{\cal H}} \stackrel{\rm LO}{=} \Re{\rm e}{\cal H}$, induced by $\Im{\rm m}{\cal H}$.
Utilizing Goloskokov-Kroll model
for $H$ GPDs \cite{Goloskokov:2006hr}, we illustrate this effect
in Fig.~\ref{fig:NLO} for $10^{-4}\le  \xi \le 10^{-2}$, accessible in a suggested Electron-Ion-Collider \cite{Boer:2011fh}, and $t=0$.  We plot $\Re{\rm e}{\cal H}$ vs.~$\xi$, for LO DVCS or TCS (solid), NLO DVCS (dashed) and NLO TCS (dotted) at
the input scale $\mu^2={\cal Q}^2 = 4 \textrm{~GeV}^2$. In the case of NLO TCS $-\Re{\rm e}{^T{\cal H}}$ is shown, since even the sign changes. We read off that the NLO correction to $\Re{\rm e}{^T{\cal H}}$ is of the order of $-400\%$ and so
the real part in TCS becomes of similar importance as the imaginary part. This NLO prediction is testable via a
lepton-pair angle asymmetry, governed by  $\Re{\rm e}{^T{\cal H}}$ \cite{TCS}. 
\section{Ultraperipheral collisions}
As described in \cite{BeKlNy} the cross section for photoproduction in hadron collisions is given by:
\begin{equation}
\sigma_{pp}= 2 \int \frac{dn(k)}{dk} \sigma_{\gamma p}(k)dk
\end{equation}
where $\sigma_{\gamma p} (k)$ is the cross section for the 
$\gamma \,p \to p\, l^+ l^-$ process and $k$ is the photon energy. 
$\frac{dn(k)}{dk}$ is an equivalent photon flux (the number of photons with energy $k$).
In Ref. \cite{PSW1} we analized the possibility to measure TCS at the LHC. The pure Bethe - Heitler contribution to $\sigma_{p p}$, integrated over  $\theta = [\pi/4,3\pi/4]$, $\phi = [0,2\pi]$, $t =[-0.05 \GeV^2,-0.25 \GeV^2]$, ${Q'}^2 =[4.5 \GeV^2,5.5 \GeV^2]$, and photon energies $k =[20,900]\GeV $  gives $\sigma_{pp}^{BH} = 2.9 \pb $. The Compton contribution (calculated with NLO GRVGJR2008 PDFs, and $\mu_F^2 = 5 \GeV^2$) gives $\sigma_{pp}^{TCS} = 1.9 \pb$.

We have choosen the range of photon energies in accordance with expected capabilities to tag photon energies
at the LHC. This amounts to a large rate of order of $10^5$ events/year at the LHC with its nominal 
luminosity ($10^{34}\,$cm$^{-2}$s$^{-1}$). The rate remains sizeable for the lower luminosity which has been achieved in 2011.
\begin{figure}[htb]
  \centering
  \includegraphics[width=0.35\textwidth]{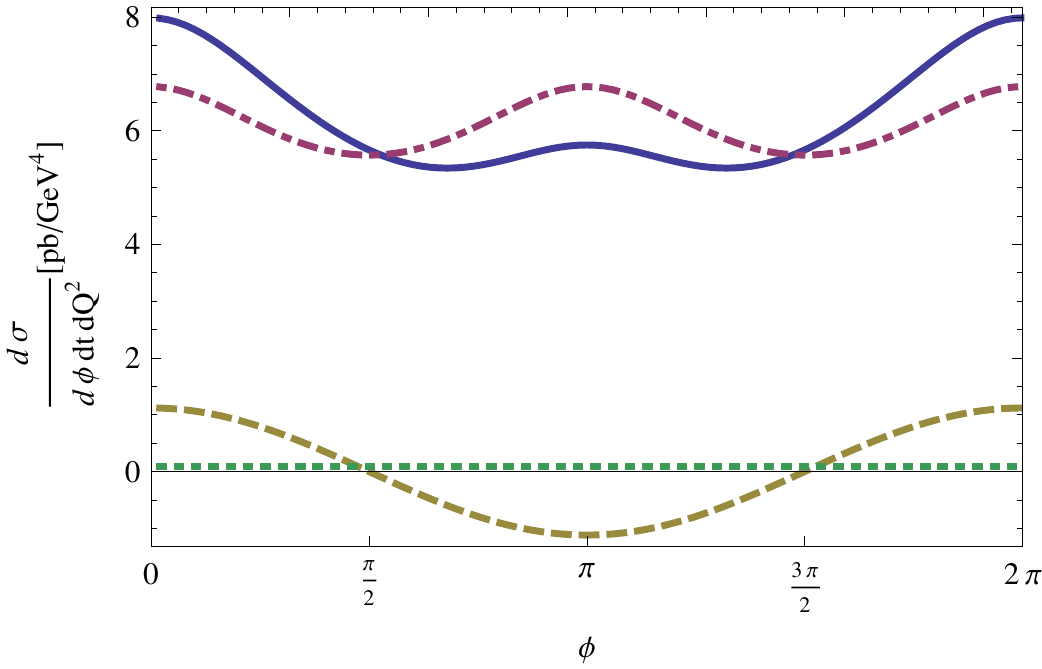}
  \caption{Total (solid), TCS (dotted), BH (dash-dotted) and intereference (dashed) differential cross section for ultraperipheral collisions at RHIC.}
  \label{Fig:RHIC}
\end{figure}
Figure \ref{Fig:RHIC} shows predictions obtained for ultraperipheral collisions at RHIC, using KG model for $t=-0.1 \GeV^2$ and $\sqrt{s_{pp}}=500 \GeV^2$. Only BH contribution gives $10^3$ events for $10^7 s$.

\section*{Acknowledgements}
This work is partly supported by the Polish Grant NCN No DEC-2011/01/D/ST2/02069 and the Joint Research Activity "Study of Strongly Interacting Matter" (acronym HadronPhysics3, Grant Agreement n.283286) under the
Seventh Framework Programme of the European Community.

{\raggedright
\begin{footnotesize}



\end{footnotesize}
}


\end{document}